\documentclass[aps,prl,twocolumn,nopacs,superscriptaddress]{revtex4}

\usepackage{graphicx}
\usepackage{color}
\usepackage{amsfonts}
\begin{document}

\title{Robust Surface Doping of Bi$_2$Se$_3$ by Rb Intercalation}

\author{Marco~Bianchi}
\author{Richard~C.~Hatch}
\author{Zheshen~Li}
\author{Philip Hofmann}\email{philip@phys.au.dk}
\affiliation{Department of Physics and Astronomy, Interdisciplinary Nanoscience Center, Aarhus University, 8000 Aarhus C, Denmark}

\author{Fei~Song}
\affiliation{Zernike Institute of Advanced Materials, University of Groningen, 9747 AG, The Netherlands and Dept. of Physics, Norwegian University of Science and Technology (NTNU), Trondheim, Norway}

\author{Jianli~Mi}
\author{Bo~B.~Iversen}
\affiliation{Center for Materials Crystallography, Department of Chemistry, Interdisciplinary Nanoscience Center, Aarhus University, 8000 Aarhus C, Denmark}

\author{Zakaria~M.~Abd~El-Fattah}
\affiliation{Centro de F\'{\i}sica de Materiales CSIC/UPV-EHU-Materials Physics Center and Donostia International Physics Center (DIPC), Manuel Lardizabal 5, 20018 San Sebasti\'an, Spain}

\author{Peter~L\"optien}
\author{Lihui ~Zhou}
\author{Alexander~A.~Khajetoorians}
\author{Jens~Wiebe}
\author{Roland~Wiesendanger}
\affiliation{Institute of Applied Physics, Hamburg University, Jungiusstrasse 11, 20355 Hamburg, Germany}

\author{Justin~W.~Wells}
\affiliation{MAX IV laboratory, Lund University, Sweden and Dept.\ of Physics, Norwegian University of Science and Technology (NTNU), Trondheim, Norway.}


\date{\today}
\begin{abstract}
Rubidium adsorption on the surface of the topological insulator  Bi$_2$Se$_3$ is found to induce a strong downward band bending, leading to the appearance of a quantum-confined two dimensional electron gas states (2DEGs) in the conduction band. The 2DEGs shows a  strong  Rashba-type spin-orbit splitting, and it has previously been pointed out that this has relevance to nano-scale spintronics devices. The adsorption of Rb atoms, on the other hand, renders the surface very reactive and exposure to oxygen leads to a rapid degrading of the 2DEGs. We show that intercalating the Rb atoms, presumably into the van der Waals gaps in the quintuple layer structure of Bi$_2$Se$_3$,  drastically reduces the surface reactivity while not affecting the promising electronic structure. 
The intercalation process is observed above room temperature and accelerated with increasing initial Rb coverage, an effect that is ascribed to the Coulomb interaction between the charged Rb ions. Coulomb repulsion is also thought to be responsible for a uniform distribution of Rb on the surface.
\end{abstract}


\maketitle
\section{Introduction}

The topologically guaranteed existence of metallic surface states on certain bulk insulators has recently attracted much interest in condensed matter physics \cite{Qi:2011,Hasan:2010}. While most research activities focus on the exotic properties of the topological surface states, the study of the materials in question has also shown them to have highly promising properties for conventional spintronics applications. Most importantly, strong band bending near the surface of the prototypical topological insulator Bi$_2$Se$_3$ \cite{Noh:2008,Hsieh:2009c}, was shown to lead to quantized two-dimensional electron gas  states (2DEGs) \cite{Bianchi:2010b} with a very strong Rashba splitting \cite{King:2011} - something which could be used to realise  nano-scale spin field effect transistors \cite{Datta:1990}. The required band bending can be induced by the controlled adsorption of a variety of species, such as CO \cite{Bianchi:2011}, H$_2$O \cite{Benia:2011}, rare earth and alkali metals, \cite{King:2011,Zhu:2011c,Valla:2012} but it is very sensitive to the amount of adsorbates, leading to an unstable electronic situation. A stabilisation can be achieved for an adsorbate coverage that is sufficiently high to saturate the band bending process \cite{Bianchi:2011,Zhu:2011c} but the particularly effective adsorption of alkali atoms can be expected to render the surface very reactive, even more than for the clean surfaces where this is already a problem  \cite{Kong:2011b,Lang:2011}. 

 Here we explore the possibility of intercalating adsorbed Rb atoms below the surface of Bi$_2$Se$_3$ and study what effect this has on the surface electronic structure and the surface stability. Compared with the situation of Rb adsorbates on the surface, we find that the intercalation of Rb atoms has almost no effect on the near-surface electronic structure as measured by angle-resolved photoemission spectroscopy (ARPES): the 2DEGs are already formed when the Rb atoms are adsorbed on the surface and they are not severely changed by the intercalation of Rb below the surface. Intercalating the highly reactive Rb atoms, on the other hand, protects them from interaction with the residual gas and the surface electronic structure is not even affected by a massive exposure to O$_2$. Real space studies by scanning tunnelling microscopy (STM) show that one of the factors driving the intercalation process is the Coulomb repulsion between the Rb atoms. This leads to a stronger tendency for intercalation at higher surface coverage and to a uniform distribution of the Rb dopants. Such control of the dopant uniformity is advantageous if spin-transport in these systems is to be exploited.

\section{Results and Discussion}

\begin{figure}
\includegraphics[width=1\columnwidth]{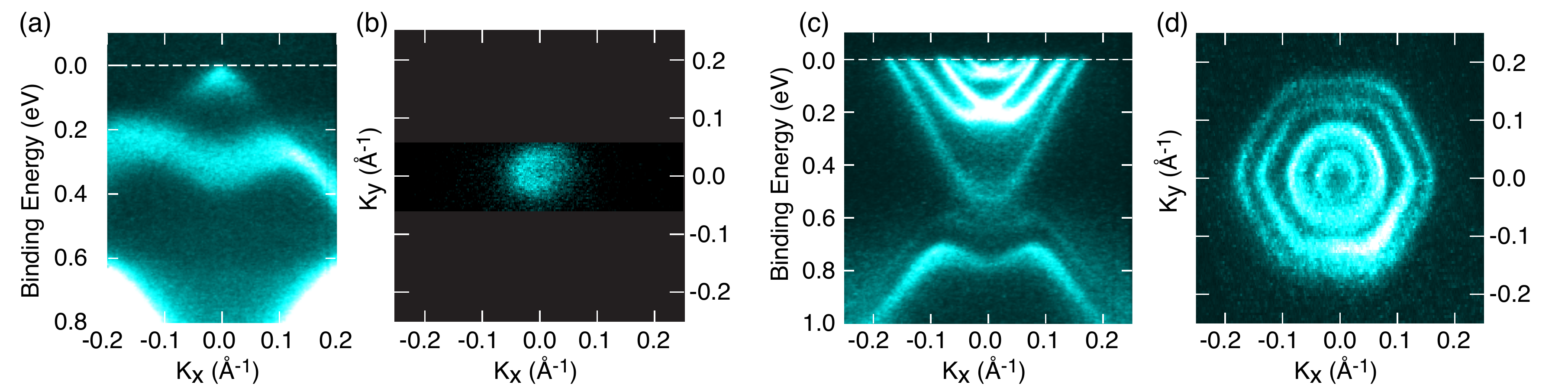}
\caption{(a) ARPES measurement of the near-surface electronic structure of clean, calcium-doped Bi$_2$Se$_3$ through the Brillouin zone center and (b) the photoemission intensity at the Fermi level (high photoemission intensity is bright). (c) The corresponding electronic structure and (d) Fermi surface after deposition of sub-monolayer Rb. Measurements are performed using  photon energies of 16~eV and 32~eV for (a),(b) and (c),(d), respectively.}\label{fermiRb}
\end{figure}

\ref{fermiRb} shows the electronic structure of calcium-doped Bi$_2$Se$_3$ after cleaving and upon the adsorption of a sub-monolayer coverage of Rb at 190~K. Initially, the topological surface state with its characteristic Dirac cone is observed to be very close to the Fermi level, resulting in a small circular Fermi surface. After Rb deposition, two Rashba-split 2DEGs can be observed in the conduction band region, leading to a complex Fermi contour consisting of five concentric features, four from the 2DEGs and one from the topological surface state (two contours nearly coincide), consistent with previous observations \cite{Valla:2012,Zhu:2011c}. The center contour is almost circular but a progressive hexagonal warping is observed for the contours further away from the center, as expected from the bulk band structure  \cite{Fu:2009,Kuroda:2010}. At the temperature of dosing (190~K) and measurement (70~K), it appears unlikely that the Rb atoms are intercalated below the surface but in the following we show that such an intercalation can be achieved by annealing the sample to higher temperatures, subsequent to low-temperature deposition. 

\begin{figure}
\includegraphics[width=\columnwidth]{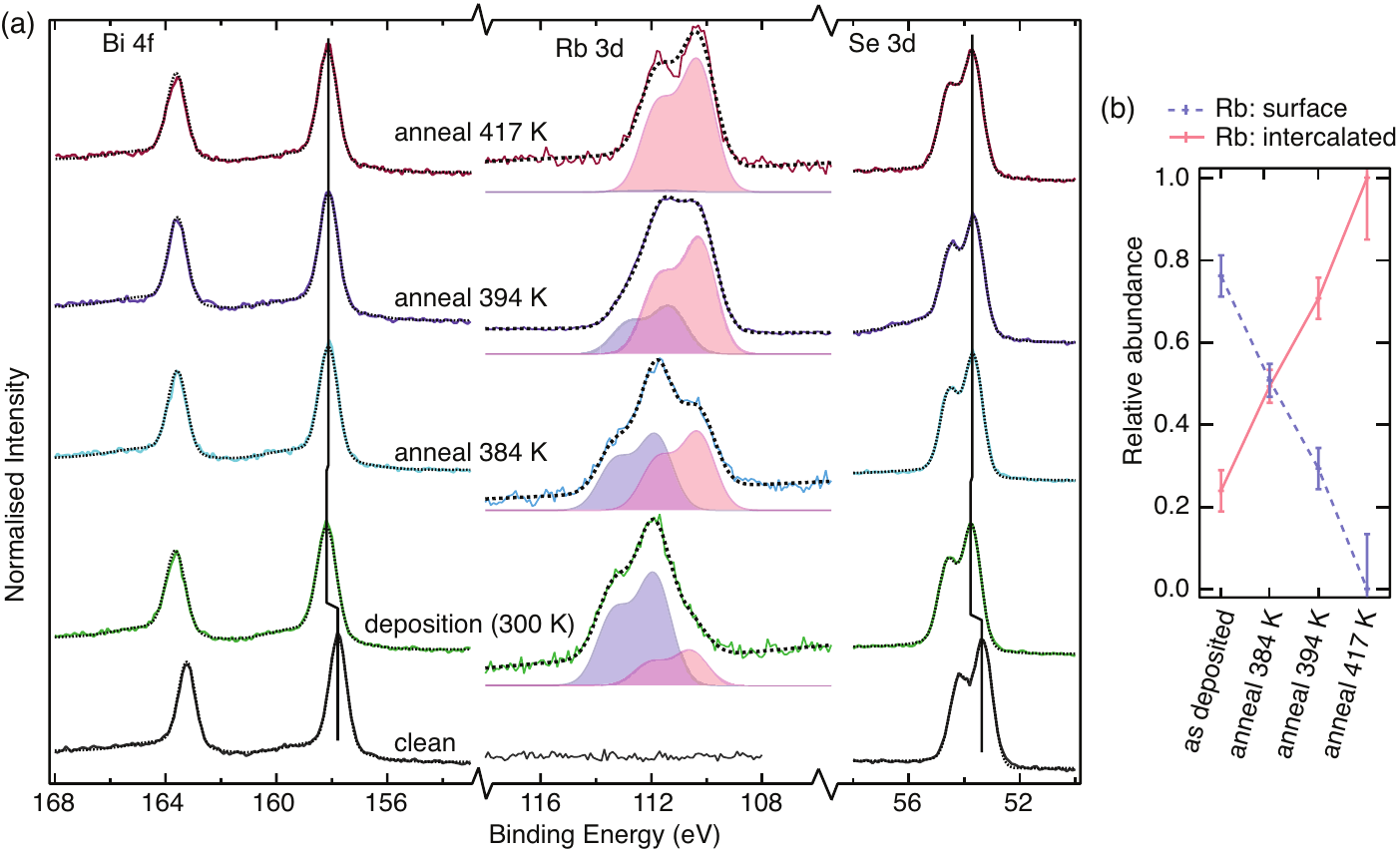}
\caption{(a) Bi 4f, Se 3d and Rb 3d core levels (coloured lines) and fits (black dotted lines) of a clean surface, following  deposition of 0.23~ML Rb and incremental annealing for 5~minutes at the temperatures given.  The Rb 3d core levels can be fitted by two doublets that are assigned to on-surface Rb (purple) and intercalated Rb (pink). Measurements are collected at room temperature using a photon energy of 353~eV and have been normalised such that the principal components are of equal intensity.  (b) Relative intensity of the surface and intercalated Rb components at selected stages of the experiment.}\label{corelevels}
\end{figure}

The temperature-induced intercalation of  Rb atoms can be followed by monitoring the Bi 4f, Rb 3d and Se 3d core levels. The bottom spectra in \ref{corelevels} show the photoemission intensity in the appropriate binding energy regions prior to Rb adsorption. The Bi 4f \cite{core_note} and Se 3d core levels are each found to be well described by two doublets; a very small minority component which can be ascribed to surface adatoms, clusters or bound impurities and a majority component assigned to atoms in the bulk of Bi$_2$Se$_3$. No surface core level shift for the outermost Se atoms is observed. The curve fits (black dotted lines) are superposed on the data for both core levels (solid coloured lines). The intensity ratio and separation is fixed following literature values \cite{Fuggle:1980a,Cardona:1978}. In each case, the position of the largest component has been highlighted by a black vertical line. 

Once $0.23$~monolayers (ML) of Rb are adsorbed at room temperature, a clear Rb 3d core level is also observable. It can also be well described by two doublets (also shown in \ref{corelevels}), each with the expected separation and intensity ratio. The intensity of the low binding energy doublet (outlined in pink) is very weak. This component will subsequently be shown to arise from the intercalated Rb atoms and its presence immediately after adsorption suggests that a degree of intercalation occurs already at room temperature.  It is also apparent that the adsorption of Rb immediately leads to a shift of the Bi and Se core level lines by 0.4~eV to higher binding energy. This is the expected behaviour due to surface doping, consistent with a similar shift of the valence band features. Due to the electrostatic nature of the shift, both the Bi and Se core level lines change their position by the same amount.  Except for this shift, the shape of these core levels is essentially unchanged. 

Annealing the sample to increasingly higher temperatures leads to the following changes in the spectra: the low binding energy doublet in the Rb 3d peak strongly increases at the expense of the high binding energy doublet (see  \ref{corelevels}(b)). We interpret this as being caused by the conversion of on-surface Rb atoms to Rb atoms intercalated below the surface, most likely in the van der Waals gaps between quintuple layers. After annealing to 417~K, the high binding energy doublet has almost vanished, suggesting that the majority of Rb atoms are now found below the surface. The relative intensity of the Rb 3d core level with respect to Bi 4f and Se 3d also decreases during the annealing.  This decrease, as well as a higher sensitivity to on-surface Rb atoms observed in grazing emission measurements,  is consistent with Rb intercalation to depths which are significant relative to the electron mean free path of $\approx$~6~\AA. A final change during annealing is a small reversal of the initial doping: the Bi 4f and Se 3d peaks move slightly back to lower binding energies, as seen by following the solid lines that mark the position of the highest peak. 

\begin{figure}
\includegraphics[width=1.0\columnwidth]{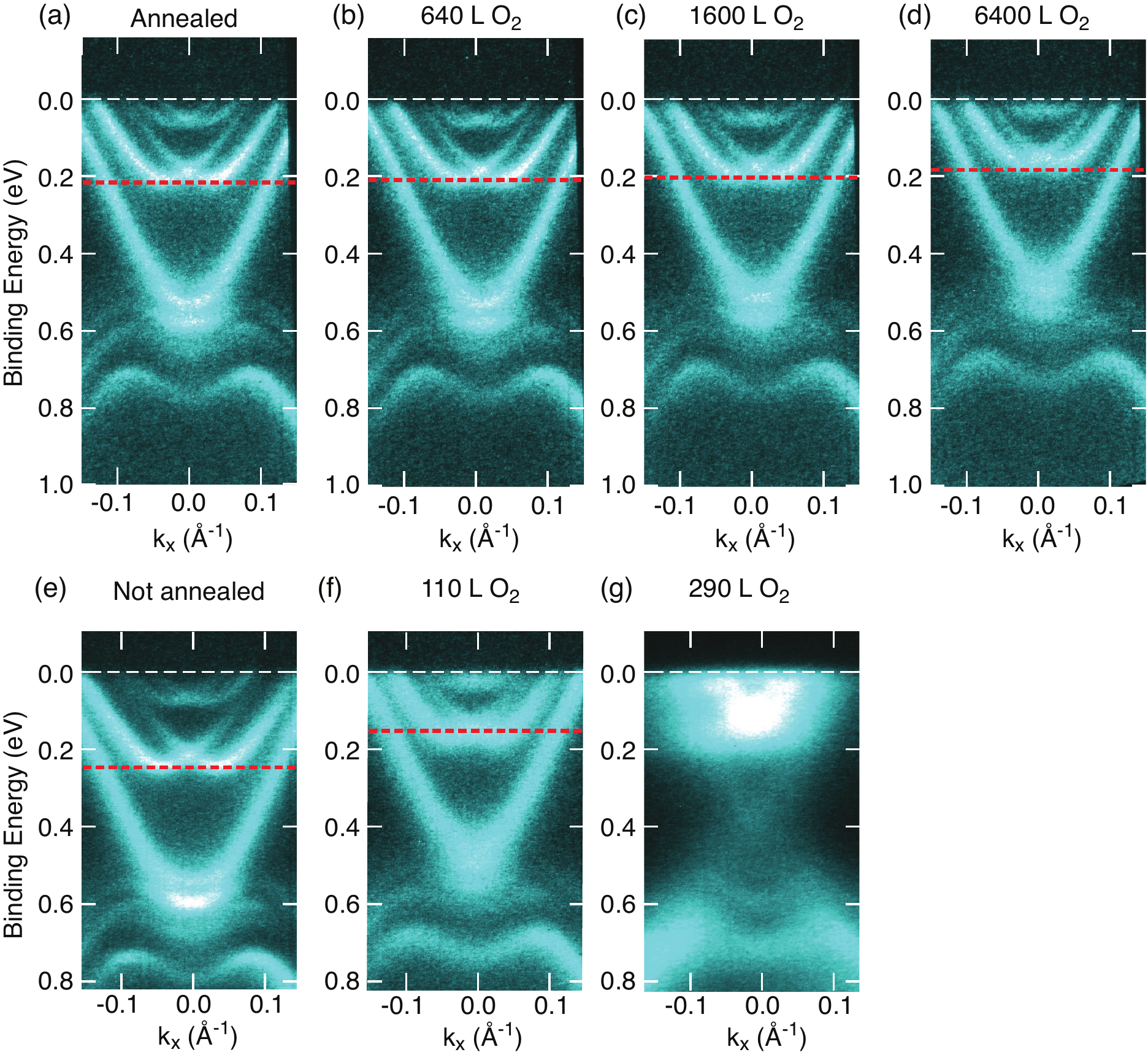}
\caption{ARPES measurements of Ca-doped Bi$_2$Se$_3$ (a) after deposition of 0.23~ML Rb at 190~K and annealing for one minute to 350~K and (b), (c) and (d) following exposure to 640~L, 1600~L and 6400~L of O$_2$, respectively. (e) A similar sample after a similar deposition of Rb at 190~K, but without subsequent annealing and (f) and (g), the same not-annealed sample following exposure to 110~L and 290~L of O$_2$, respectively. Data are collected at 70~K using a photon energy of 16~eV. In each case, a red dotted line indicates the approximate position of the first Rashba-split 2DEG, such that a comparison of the extent of doping is easier.}\label{slices}
\end{figure}

Although the effect of Rb intercalation is very pronounced in the core levels, it hardly affects the electronic structure in the valence and conduction bands.  This can be seen by comparing  \ref{slices}(a) with \ref{slices}(e) where a similar amount of Rb has been dosed at 190~K, but the sample in \ref{slices}(a) has been annealed (in this case to 350~K)  whereas the sample in  \ref{slices}(e) has not. In the annealed case, the electronic structure is slightly sharper, and the doping strength is slightly less, but otherwise the two cases are remarkably similar. This is significant since the location of the Rb is entirely different in these two cases, yet the electronic structure and surface doping are essentially unaffected. 

While the annealing and intercalation hardly affects the electronic structure of the surface, \ref{slices} illustrates that the annealed surface is much more stable towards a reaction with oxygen: after 6400~L of O$_2$ exposure, the doping is marginally reduced to $\approx$0.60~eV, but neither the 2DEGs nor the topological surface state are significantly degraded. Conversely, a similar experiment in which Rb is dosed at 190~K  without further annealing reveals that the surface is quickly degenerated by exposure to O$_2$; already after $\approx$300~L, the conduction band quantum well states are barely discernible. 

\begin{figure}
\includegraphics[width=1.0\columnwidth]{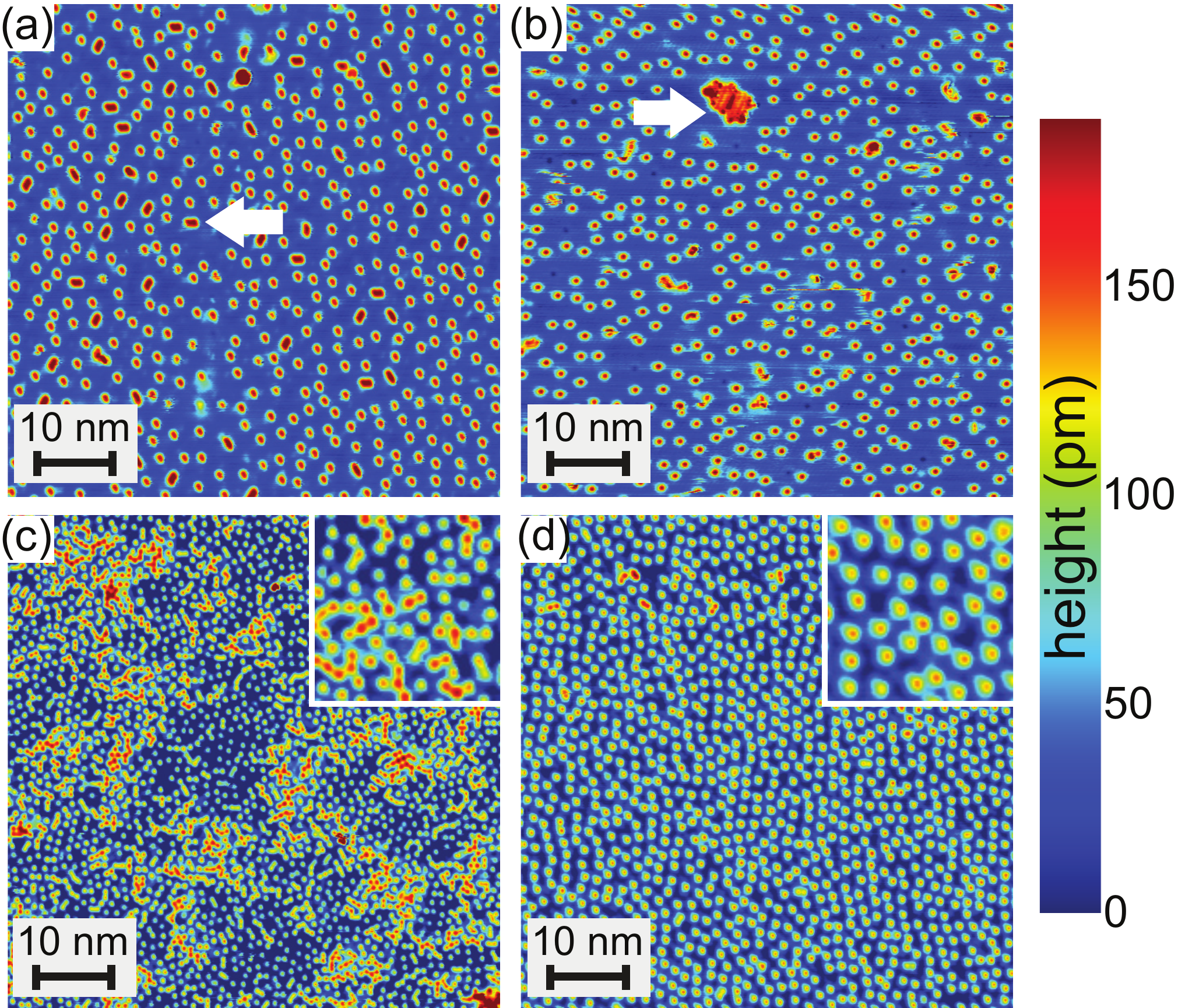}
\caption{(a,b) STM topograph of 0.025~ML Rb on Ca-doped Bi$_2$Se$_3$ before (a) and after (b) annealing at 400~K for 10~min. After annealing the on-surface Rb coverage is reduced by 20\% (b). Arrows: see text. (c,d) Corresponding data for an initial coverage of 0.12~ML Rb before (c) and after (d) annealing. After annealing the on-surface Rb coverage is reduced by 60\% (d). The insets (10~nm$\times$10~nm) show magnified views of the samples. (Tunneling parameters: $V =$1~V, $I =$15~pA, measurement temperature: 1.2~K.) }\label{stm_intercal}
\end{figure}

The intercalation process and its origin can be further studied by STM.
\ref{stm_intercal}(a)-(d) show STM topographs of samples with two different Rb coverages before and after annealing at 400~K for 10~min. In all samples the major features are single, well separated Rb atoms, which appear as circular protrusions with a height of $\approx$150~pm. They are thus easily discernible from the trigonal features of  Se vacancies or from sub-surface Ca dopants.\cite{Hor:2009}. In fact, the corrugation of the latter features is more than ten times smaller so that they can only be observed in images taken for a substantially lower Rb coverage. For the sample covered with 0.025~ML Rb (a,b), a few dimers are visible before annealing [see arrow in (a)] and after annealing some clusters are formed, most probably from residual gas adsorption [see arrow in (b)]. After annealing, the coverage of on-surface Rb atoms is reduced by 20\% and the dimers have disappeared. In contrast, for the sample covered with 0.12~ML Rb (\ref{stm_intercal} c,d), which is closer to the coverage used in the ARPES and core level photoemission measurements, there is a much stronger effect of the annealing on the distribution and coverage of on-surface Rb atoms. \ref{stm_intercal}(c) shows the STM topograph of the non-annealed sample after Rb adsorption. We find a distribution of well separated Rb atoms with an increase in the apparent height in areas of locally higher coverage, probably due to charge accumulation effects \cite{Wiebe:2003,Beidenkopf:2011}. After annealing, the coverage of on-surface Rb atoms is strongly reduced to 0.05~ML (\ref{stm_intercal}(d)) indicating that 60\% of Rb has diffused into the bulk. Moreover, the distribution of the Rb remaining adsorbed on the surface is much more homogeneous after annealing. The enhanced tendency for intercalation at high coverage, as well as the trend for a more uniform distribution of  the Rb atoms after annealing, can both be ascribed to the strong Coulomb interaction between the Rb atoms that are highly ionic due to a charge transfer to the bulk \cite{Valla:2012,Song:2012}. 

Our results show that the annealing of the  Bi$_2$Se$_3$ sample after adsorption of Rb leads to a decrease of Rb coverage on the surface. While we do not directly show that the disappearing Rb atoms are intercalated into the van der Waals gaps between the quintuple layers making up the  Bi$_2$Se$_3$ structure, this appears very likely. In fact, such intercalation processes have been known for a long time \cite{Paraskevopoulos:1988} and intercalation can be achieved during bulk crystal growth  \cite{Hor:2010b} or using electrochemical methods \cite{Kriener:2011}. 

Disappearing on-surface Rb atoms after annealing could be explained by thermal desorption instead of intercalation, as has been proposed in the case of potassium doping.\cite{Zhu:2011c} However, this alternative mechanism cannot be reconciled with the other experimental findings in the present case. Most importantly, a thermal desorption of the Rb atoms could be expected to reverse the observed band bending, such that the initial electronic structure is recovered, something that clearly is not the case. Desorption would also lead to a mere decrease of the Rb 3d core level intensity, not to the appearance of a new component in the spectrum. Finally, the desorption of alkali atoms from metal \cite{Albano:1986} or semiconductor \cite{Tanaka:1990} surfaces typically takes place at much higher temperatures. Since thermal desorption appears not to play a role here it is somewhat surprising that it should for the chemically similar K. Further direct structural investigations, for example by STM, should be able to clarify this point.  An interesting similarity between thermal desorption and the intercalation observed here is the fact that higher alkali coverages favour both processes. In the case of desorption, a higher coverage lowers the desorption temperature, here it lowers the temperature required to achieve intercalation. In thermal desorption, this effect can be described by a simple model that takes the Coulomb interaction between the adsorbates into account \cite{Albano:1986}.

So far, the intercalation of dopants from surface adsorbates has only been reported for the case of Ag on Bi$_2$Se$_3$ \cite{Ye:2011}, where this is observed already at room temperature. Also in that case, the Ag adsorption leads to a shift of the entire electronic structure to higher binding energies and to the appearance of additional states in the conduction band region. 
This, however, is not interpreted as band bending but ascribed to an increased spacing of the van der Waals gap, caused by the intercalation of Ag. Indeed, first principles calculations can bring about features similar to the experimental ones, but only if an increase of the van der Waals gap by 36$\%$ is assumed. A similar increase of the van der Waals gap had also been evoked previously to explain not only the Rashba-split 2DEGs in the conduction band but also the M-shaped features in the valence band \cite{Eremeev:2011b} that emerge after the adsorption of Rb  (see \ref{slices}).

Band bending, on the other hand, can not only explain the formation of the Rashba-split 2DEG in the conduction band region but also the M-shaped states in the valence band. These can also be described as quantum well states caused by an overall downward bending of the bands. In the case of valence band states, this is unusual because only upward band bending would be expected to lead to quantized hole states. In the special case of Bi$_2$Se$_3$, however, it can also quantize the valence band because the upper valence band is rather narrow and placed over a projected bulk band gap \cite{Bianchi:2011}. In this picture, a full quantization of the valence band states would require a band bending larger than the total bandwidth ($\approx$~300~meV) but the lowest lying states could already be quantized for a smaller band bending.

Evidently, Rb adsorption has a very pronounced effect on the near-surface electronic structure but the formation of 2DEGs in the conduction band and of the M-shaped states in valence band region already takes place when Rb atoms are adsorbed on the surface. The further intercalation brings about only very minor additional changes. 
On the other hand, an increased van der Waals interlayer spacing upon intercalation is very likely but it is probably quite small. In fact, bulk x-ray diffraction measurements have found such an increase for Cu-intercalated Bi$_2$Se$_3$ to be smaller than 3$\%$ \cite{Hor:2010b}. 

Another important result  is that the intercalation of Rb not only increases the chemical stability, for example towards oxidation. The saturated strong band bending also results in a very stable situation towards further changes in the electronic structure by the adsorption of other species that otherwise increase the band bending. The electronic structure of clean  Bi$_2$Se$_3$ is very sensitive to the adsorption of minute quantities of residual gas in ultra-high vacuum (UHV) \cite{Hatch:2011} because this causes a band bending. The adsorption of Rb improves this situation because the band bending can be saturated, similar to the situation recently reported  for K adsorption at low temperature \cite{Zhu:2011c}. Exposing the non-intercalated system to UHV for 24 hours leads to a reduction of the band bending by $\approx$~100~meV. The full chemical protection is achieved by the intercalation of Rb that results in a surface that remains stable in UHV overs several days.

\section{Conclusion}
We have demonstrated that Rb atoms adsorbed on Bi$_2$Se$_3$ can be intercalated below the surface by annealing above room temperature. This results in a very stable situation with respect to further band bending and chemical reactivity. Intercalation of adsorbates is a promising way to influence the electronic structures of bulk topological insulators where it is most relevant: near the surface. Controlled intercalation processes could be an important step in the fabrication of topological insulator-based devices before further surface passivation is carried out \cite{Lang:2011}. In particular, the intercalation of species giving rise to an upward band bending could lead to a space-charge layer that is depleted of electrons, even if the bulk material is strongly $n$-doped, as often the case for these materials. 

\section{Experimental Details}

Bulk Bi$_2$Se$_3$ crystals were calcium doped to position the bulk Fermi level within the bulk band gap. 
These samples were synthesized by melting a mixture of Bi (5N purity) , Se (5N purity) and Ca (99.5\% purity) with ratio Bi : Se : Ca = 1.996 : 3 : 0.004 at 860$^{\circ}$C for 24 hours in an evacuated quartz ampoule. Once cooled down from 860$^{\circ}$C to 750$^{\circ}$C at a rate of 50$^{\circ}$C/h and then from 750$^{\circ}$C to 600$^{\circ}$C at a rate of 2~$^{\circ}$C/h, the sample was annealed at 600$^{\circ}$C for 7 days. 

ARPES and core level photoemission measurements were performed at beamlines SGM III \cite{Hoffmann:2004} and SGM I, respectively, at the synchrotron radiation source ASTRID. Clean  Bi$_2$Se$_3$ surfaces were obtained by cleaving bulk crystals in UHV, and the surface state Dirac point was measured to be less than 50~meV below the Fermi level. Exposure to Rb and O$_2$  as well as annealing were performed \textit{in situ}. The surface coverage of Rb was determined by the relative intensity of Rb and substrate core level lines. It is given in monolayers (ML), \emph{i.e.} relative to the number of Se atoms in the first layer. The surface was exposed to molecular oxygen and the exposure is given in Langmuir (L). Rb was typically dosed within 20 minutes of cleaving, such that surface doping due to ageing can be considered insignificant.

We have also carried out ARPES experiments for Rb adsorption on pristine Bi$_2$Se$_3$ samples to test the the role of the Ca doping. As expected due to the small quantity of Ca, we have found no indication of different behavior. Annealing the Rb-dosed samples, one can achieve a smaller but equally stable band bending and a surface that is stable in UHV over several days.

The STM experiments were performed in an UHV low-temperature scanning tunneling microscope facility cooled by a Joule-Thomson cryostat \cite{specs_stm}. The bulk Bi$_2$Se$_3$ samples were cleaved in UHV at room temperature and immediately transferred into the cryogenic microscope. During cool down, Rb was deposited onto the substrates at T < 200~K. For annealing, the samples were transferred in UHV from the STM to a Boron Nitride heater with a thermocouple. STM topographs were acquired at 1.2~K in the constant current mode at a current $I$ with a bias voltage $V$ applied to the sample. The coverage of on-surface Rb was determined by counting the number of Rb atoms visible in STM topographs of a certain area and dividing by the number of surface Se unit cells within the same area. We have also performed annealing experiments for the bare calcium-doped samples. For annealing temperatures below 423~K, we did not notice segregation of Se or Ca to the surface.

The use of three separate experimental systems means that there can be a systematic uncertainty between the temperatures quoted, which we estimate could deviate by up to 20~K from one system to another. It also means that experimental parameters such as Rb dose, and annealing and measuring temperatures, are not identical. Since similar behaviour is observed over an extended parameter range, this is not thought to be significant. Also, the Rb coverages are determined with a rather different accuracy in the STM and the photoemission setups. The coverage determination by STM is highly accurate but the uncertainty in the photoemission measurements can be  larger, and may be up to 20\% for the quoted values.

This work was supported by a grant from the Lundbeck foundation. 
J. W. Wells acknowledges financial support from the Institute for Storage Ring Facilities, Aarhus University. Z. M. Abd-el-Fattah acknowledges both Spanish (MAT2010-21156-C03-01) and Basque Government (IT-257-07) for funding. P. L\"optien, L. Zhou, A. A. Khajetoorians, J. Wiebe, and R. Wiesendanger acknowledge financial support from the Free and Hanseatic City of Hamburg \emph{via} the Cluster of Excellence NANO-SPINTRONICS, from the Deutsche Forschungsgemeinschaft \emph{via} the Graduiertenkolleg ``Functional Metal-Semiconductor Hybrid Systems'', and from the European Research Council (ERC) Advanced Grant ``FURORE''.
%

\end{document}